\documentclass[pageno]{jpaper}

\usepackage[normalem]{ulem}
\usepackage{authblk}
\usepackage{hyperref}
\hypersetup{colorlinks=true, urlcolor=black}
\usepackage[labelfont=bf,textfont=md]{caption}
\usepackage{subfig}
\usepackage{graphicx}
\usepackage{makecell}
\usepackage{multirow}
\usepackage{tablefootnote}

\usepackage{tikz}

\begin{document}

%\title{Hardware Mechanisms to Support Multi-Task Execution on a CGRA Using Flexible Hardware Resource Partitioning and Dynamic Partial Reconfiguration}
\title{Hardware Abstractions and Hardware Mechanisms to Support Multi-Task Execution on Coarse-Grained Reconfigurable Arrays}
\author{Taeyoung Kong}
\author{Kalhan Koul}
\author{Priyanka Raina}
\author{Mark Horowitz}
\author{Christopher Torng}
\affil{Stanford University\\
\href{kongty@stanford.edu}{\url{{kongty, kkoul, praina, horowitz, ctorng}@stanford.edu}}}
\date{}
\maketitle

\thispagestyle{empty}

%%%%%% -- PAPER CONTENT STARTS-- %%%%%%%%

\begin{abstract}
Domain-specific accelerators are used in various computing systems ranging from edge devices to data centers.
Coarse-grained reconfigurable arrays (CGRAs) represent an architectural midpoint between the flexibility of an FPGA and the efficiency of an ASIC and are a promising candidate for servicing multi-tasked workloads within an application domain.
Unfortunately, scheduling multiple tasks onto a CGRA is challenging. CGRAs lack abstractions that capture hardware resources, leaving workload schedulers unable to reason about performance, energy, and utilization for different schedules.
This work first proposes a CGRA architecture that can flexibly partition key resources, including the global buffer memory capacity, the global buffer memory bandwidth, and the compute resources.
Partitioned resources serve as hardware abstractions that decouple compilation and resource allocation.
The compiler uses these abstractions for coarse-grained resource mapping, and the scheduler uses them for flexible resource allocation at run time.
We then propose two hardware mechanisms to support multi-task execution.
A flexible-shape execution region increases the overall resource utilization by mapping multiple tasks with different resource requirements.
Dynamic partial reconfiguration (DPR) enables a CGRA to update the hardware configuration as the scheduler makes decisions rapidly.
We show that our abstraction can help automatic and efficient scheduling of multi-tasked workloads onto our target CGRA with high utilization, resulting in
1.05x--1.24x higher throughput and a 23--28\% lower % turn-around time
latency in a multi-tasked cloud workload and 60.8\% reduced latency in an autonomous system workload when compared to a baseline CGRA running single tasks at a time.
\end{abstract}

\section{Introduction}
\label{sec:introduction}

% Deep Neural Networks (DNNs) have been widely adopted over the last decade in diverse domains including computer vision, natural language processing, and recommendation.
% As state-of-the-art DNNs require large amount of both computation and memory usage, various types of DNN hardware accelerators have been proposed to improve performance and energy-efficiency.
% Many vendors have designed their own hardware accelerators that are specific to DNN tasks.
% While these ASIC accelerators show optimal performance and power-efficiency for the target tasks, they cannot accommodate the evolution of the applications.
% Some vendors have deployed DNNs on FPGAs as it provides flexibility to support many applications and better power-efficiency than general-purpose processors.
% However, FPGAs are expensive due to their bit-level flexibility and difficult to program.
% Coarse-grained reconfigurable arrays (CGRAs) have gained interest in recent years as they provide balance between flexibility, performance, and energy-efficiency.
% Their word-level flexibility at interconnect and processing elements (PEs) enables efficient hardware reconfigurability to accommodate many applications, which makes them serve as a midpoint between FPGAs and ASICs.

Domain-specific accelerators have gained growing interest in recent years as they provide improved performance and energy efficiency over general-purpose processors.
Application-specific integrated circuits (ASICs)~\cite{eyeriss, eie, tpu} show the highest performance and efficiency as they are specialized for target applications such as image processing or machine learning (ML).
However, the ASIC design process can span multiple years, and fixed-function accelerators quickly become obsolete as applications continue to evolve. 
Some works deploy applications on FPGAs~\cite{msftfpga, angeleye, ese}.
FPGAs enable reconfiguration of the underlying hardware and can accelerate diverse workloads, but their bit-level flexibility incurs high area and energy overheads.
Coarse-grained reconfigurable arrays (CGRAs) are promising architectures that lie between ASICs and FPGAs.
A CGRA has arithmetic units and a routing system that are configurable in word-level granularity, providing flexibility at a lower overhead than a FPGA.
With its unique advantages, a CGRA can be widely adopted in domains with high performance, power, and flexibility requirements.

As hardware accelerators are deployed in various scenarios, the demand for multi-task execution support on hardware is growing.
For example, many vendors~\cite{tpu, brainwave} offer INFerence-as-a-Service, where multiple tenants share the same hardware to run inference tasks.
Also, an autonomous system handles concurrent tasks to process various types of data from numerous sensors.
Some works have explored multi-task execution support in ASICs and FPGAs.
PREMA~\cite{prema} and Planaria~\cite{planaria} propose a systolic array that supports multi-tenancy by temporal and spatial multiplexing, respectively.
\cite{vital, fpgavirt, deepfpga} propose an FPGA virtualization framework with multi-tenancy support.
However, multi-task execution support on CGRAs has not been explored much thus far.
A noteworthy exception is ChordMap~\cite{chordmap} which schedules multiple tasks captured in synchronous data flow graphs onto a CGRA.
However, it assumes that all tasks are known a priori, whereas in a multi-tenant cloud or multi-tasked edge workload scenario, tasks may arrive dynamically and require schedulers to react to maximize utilization.

Unfortunately, scheduling multiple tasks onto a CGRA is challenging as it lacks abstractions capturing hardware resources.
In this paper, we propose hardware abstractions of a CGRA by partitioning key hardware resources.
Both compilers and schedulers can exploit the abstractions to reason about performance, energy, and utilization.
We also develop hardware mechanisms that allow fast and flexible multi-task execution on a CGRA, which schedulers exploit to improve hardware utilization.
We evaluate our CGRA with two different multi-tasked workload scenarios to show the potential.
Our key contributions are:
\begin{itemize}
   \item \raisebox{.5pt}{\textcircled{\raisebox{-.9pt} {1}}}
We propose a CGRA architecture that can flexibly re-partition key resources, including the Global Buffer (GLB) memory capacity, the GLB memory bandwidth, and the compute resources.
Specifically, we partition the GLB into GLB-slices and the tile array into array-slices, which serve as hardware abstractions.
The compiler uses these abstractions for coarse-grain resource mapping, while the scheduler uses them for flexible resource allocation.

   \item \raisebox{.5pt}{\textcircled{\raisebox{-.9pt} {2}}}
We propose two hardware mechanisms to support multi-task execution on the CGRA.
First, the CGRA can form a flexible-shape execution region at run time.
It improves resource utilization by enabling a scheduler to allocate GLB-slices and array-slices flexibly.
Second, we propose a fast-DPR method to reconfigure the underlying hardware rapidly according to scheduler decisions.
% Our fast-DPR allows reconfiguration of the entire tile array in 3.5{\micro}s without any pipeline registers.
It also supports run time relocation of a task to any available array-slice without software intervention.

   \item \raisebox{.5pt}{\textcircled{\raisebox{-.9pt} {3}}}
We quantify the benefits of our proposed mechanisms on two different examples.
Our CGRA with flexible execution regions and fast-DPR shows 1.05x--1.24x higher throughput and 23--28\% lower latency in a cloud system scenario and 60.8\% reduced latency in an autonomous system scenario than the baseline CGRA.

\end{itemize}
\section{Architectural Support for Multi-Task Execution on a CGRA}
\label{sec:hardware}

% \begin{figure}
%     \centering
%     \subfloat[SDM]{\includegraphics[width=0.47\textwidth,height=2cm]{example-image}
%     \label{fig:sub1}}\hskip1ex
%     \subfloat[TDM]{\includegraphics[width=0.47\textwidth,height=2cm]{example-image}
%     \label{fig:sub2}}
%     \caption{SDM TDM}
%     \label{fig:sdmtdm}
% \end{figure}

In this section, we explore the architectural support needed for multi-task execution on a CGRA.
Section~\ref{subsec:baseline} first introduces a baseline CGRA architecture with common features present in many reconfigurable accelerators~\cite{amber, jade, snafu, dream, egra, cgravideo}.
Section~\ref{subsec:abstraction} then introduces how we abstract the hardware resources in the CGRA for the scheduler by partitioning the global buffer (GLB) and the tile array into GLB-slices and array-slices, respectively.
We further develop hardware mechanisms that enable multi-task execution on top of these abstractions (Section~\ref{sec:partition}), including
flexible-shape execution regions and dynamic partial reconfiguration (DPR).

\subsection{Baseline CGRA Architecture}
\label{subsec:baseline}

\begin{figure}
    \centering
    \includegraphics[width=1.0\linewidth]{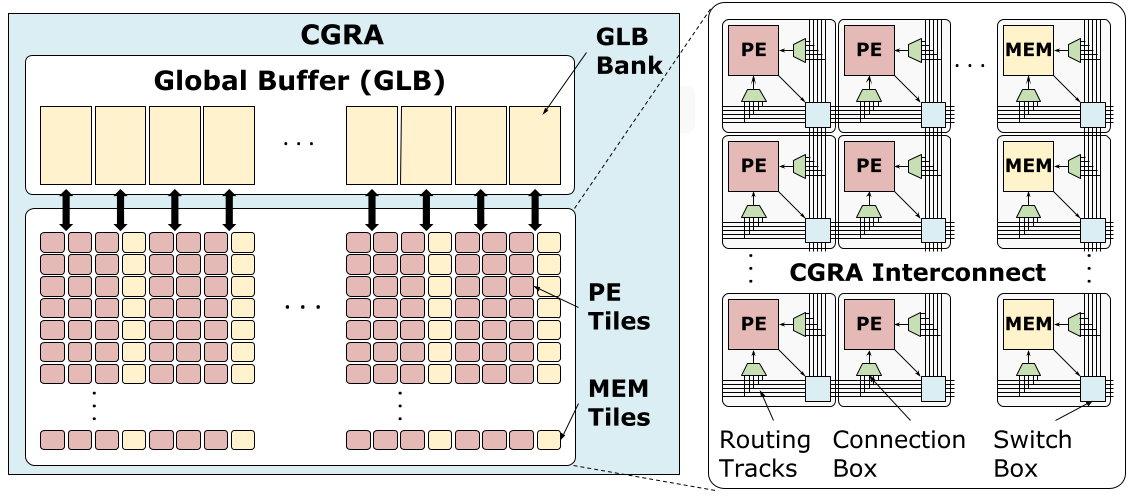}
    \caption{Baseline CGRA block diagram corresponding to~\cite{ahatecs}.}
    \label{fig:cgra-baseline}
\end{figure}

Our baseline CGRA consists of a tile array with processing element (PE) and memory (MEM) tiles and a global buffer (GLB) (Figure~\ref{fig:cgra-baseline}).
We leverage almost the same hardware configuration used in the Amber SoC~\cite{amber}.
The CGRA has 32x16 tiles with 384 PE tiles and 128 MEM tiles, and tiles communicate through a statically configured mesh interconnect.
A PE tile is extended from Amber version to support MAC operation.
Each node in the interconnect has five incoming and five outgoing tracks in each direction, and switch boxes route data from incoming tracks to outgoing tracks.
Connection boxes select data from incoming tracks and route it to the PE or MEM tile cores.
The GLB consists of 32 banks, with each bank containing 128 KB of SRAM.
Each GLB bank directly communicates with the tile array through IO tiles located at the top of the array.

\subsection{A Scheduler-Visible Abstraction of Hardware Resources}
\label{subsec:abstraction}

\begin{figure*}
    \centering
    \captionsetup[subfigure]{oneside,margin={-1cm,0cm}}
    \subfloat[Baseline]{\includegraphics[width=0.44\textwidth]{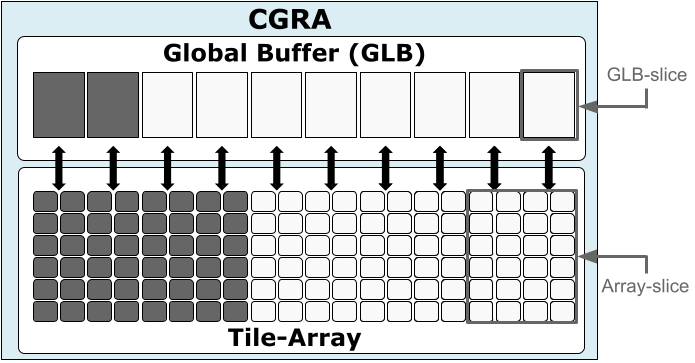}
    \label{fig:resource-baseline}}
    \subfloat[Fixed-sized execution region]{\includegraphics[width=0.44\textwidth]{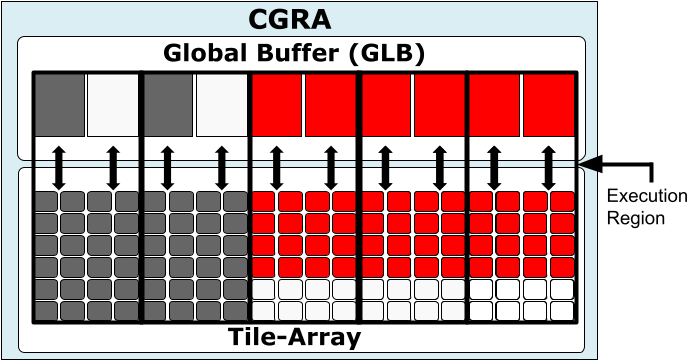}
    \label{fig:fixed}}
    \vspace{-3pt}
    \subfloat[Variably sized execution region]{\includegraphics[width=0.44\textwidth]{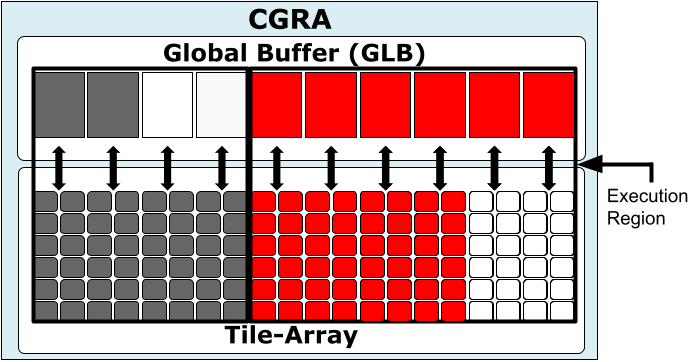}
    \label{fig:variable}}
    \subfloat[Flexible-shape execution region]{\includegraphics[width=0.44\textwidth]{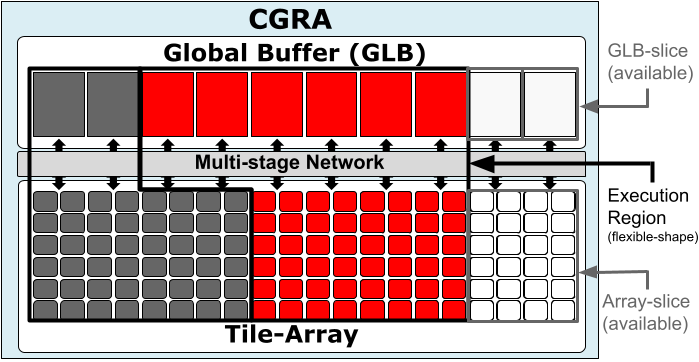}
    \label{fig:flexible}}
    \caption{Resource allocation in the baseline CGRA and a CGRA with three different execution regions. Resources colored grey represent the blocks occupied by a current-running task, and those colored red represent blocks occupied by a next-running task.}
    \label{fig:partition}
\end{figure*}

\begin{table}
\small
\centering
\begin{tabular}{ c | c | c | c | c | c} 
\toprule\toprule
App. & Task & Ver. & Tpt. &  \makecell{Array \\ slices} & \makecell{GLB \\ slices} \\  \midrule
\multirow{8}{*}{ResNet-18} & \multirow{2}{*}{conv2\_x} & a & 64 & 2 & 7  \\
                        &                           & b & 256 & 6 & 7  \\ \cline{2-6}
                        & \multirow{2}{*}{conv3\_x} & a & 64 & 2 & 4  \\
                        &                           & b & 256 & 6 & 4  \\ \cline{2-6}
                        & \multirow{2}{*}{conv4\_x} & a & 64 & 2 & 6  \\
                        &                           & b & 256 & 6 & 6  \\ \cline{2-6}
                        & \multirow{2}{*}{conv5\_x} & a & 64 & 2 & 20  \\
                        &                           & b & 128 & 6 & 20  \\ \cline{1-6}
\multirow{6}{*}{MobileNet}  & \multirow{2}{*}{\makecell{conv\_dw \\ \_pw\_2\_x}\tablefootnote{A \textit{conv\_dw\_pw} refers to a merged task of a depth-wise convolutional layer and a point-wise convolutional layer.}} & a & 52 & 2 & 4  \\
                            &                                                    & b & 208 & 5 & 4  \\ \cline{2-6}
                            & \multirow{2}{*}{\makecell{conv\_dw \\ \_pw\_3\_x}} & a & 52 & 2 & 4  \\
                            &                                                   & b & 104 & 3 & 4  \\ \cline{2-6}
                            & \multirow{2}{*}{\makecell{conv\_dw \\ \_pw\_4\_x}} & a & 52 & 2 & 4  \\
                            &                                                    & b & 104 & 3 & 4  \\ \cline{1-6}
\multirow{2}{*}{\makecell{Camera\\pipeline}}  & \multirow{2}{*}{\makecell{Camera\\pipeline}} & a & 3 & 4 & 4  \\
                                              &                                             & b & 12 & 6 & 14  \\ \cline{1-6}
\multirow{3}{*}{Harris}  & \multirow{3}{*}{Harris} & a & 1 & 2 & 4 \\ 
                         &                        & b & 2 & 4 & 7  \\ 
                        &                         & c & 4 & 7 & 14  \\ \cline{1-6}                            
\bottomrule
\end{tabular}
\caption{Variants of tasks with different resource usage and throughput. ResNet-18 and MobileNet consist of several layers, and one or more layers form a single task. The unit of throughput (Tpt.) for ResNet-18 and MobileNet is MACs/cycle and for camera pipeline and harris it is pixels/cycle.}
\label{tab:task-resource}
\end{table}

We focus on three key hardware resources within the CGRA (Figure~\ref{fig:cgra-baseline}): the GLB memory capacity, the GLB memory bandwidth, and the compute resources within the tile array. When a task is compiled in the Amber toolchain~\cite{ahatecs}, a compiler converts it into a dataflow graph where each node and edge represents a hardware resource and communication, respectively.
Specifically, GLB banks are used for medium-sized storage and communication to the host and tile array, and PE and MEM tiles are used for computation and as small scratchpads.
The dataflow graph can derive the usage of memory capacity, memory bandwidth, compute units, and throughput.

We abstract the hardware resources by partitioning the GLB and tile array into homogeneous GLB-slices and array-slices, respectively.
% , with corresponding architectural support to enable dynamic allocation to scheduled tasks.
For example, we can abstract each GLB bank within our CGRA as a GLB-slice and every set of four columns in the tile array (48 PE tiles and 16 MEM tiles) as an array-slice.
This abstraction serves as a middle layer that decouples offline bitstream generation by a compiler and run time resource allocation by a scheduler.
During compilation, we represent the resource usage of each task using these abstracted GLB-slices and array-slices.
For instance, a \textit{conv2\_x} layer in~\cite{resnet} utilizes 750KB of GLB memory capacity, 17.3MB/s of memory bandwidth, 80 PE tiles, and 17 MEM tiles and achieves 64 OPs/cycle throughput at a 500MHz clock frequency.
The task is abstracted as seven GLB-slices and two array-slices in coarse-grain resource slice usage.
It is possible to produce variants of the same task with different resource usage and throughput by tweaking the compiler.
For example, increasing the unroll factor of the same task by four would achieve 4x throughput (256 OPs/cycle) with 288 PE tiles, 33 MEM tiles, and the same GLB memory capacity and bandwidth, which is abstracted as seven GLB-slices and six array-slices.
Our approach allows for pre-computation of bitstreams that support different resource usage and throughput to be cached in on-chip storage to support fast dynamic partial reconfiguration, as discussed later.
Table~\ref{tab:task-resource} summarizes the resource usage and throughput for several different variants of tasks.
At run time, a scheduler leverages the hardware slice abstraction to decide which variant of tasks to choose, which resources to allocate, and when to execute.
% The next section describes hardware mechanisms exploit these hardware abstractions to support multi-task execution.

\subsection{Hardware Mechanisms}
\label{sec:partition}
\noindent\textbf{Flexible-Shape Execution Regions}.
% Now that we have hardware abstractions of the CGRA resources as described in Section~\ref{subsec:abstraction},
To manage multiple tasks that are concurrently running, we need a way to monitor hardware resources and the status of tasks, that are build upon the abstractions described above.
We introduce an \textit{execution region}, a sub-region of the CGRA on which a single task is mapped and executed.
An execution region consists of one or more GLB-slices and array-slices.
The flexibility to form different sizes and shapes of execution regions gives the scheduler a simplified and quantized view of hardware resources while providing enough information to allocate resources to each task to maximize resource utilization in multi-tasked workloads.

Figure~\ref{fig:partition} compares different mechanisms to form an execution region and how they affect resource allocation.
The blocks colored in gray represent resources occupied by the currently running task, and those colored in red represent resources allocated to the next-running task.
The baseline CGRA (Figure~\ref{fig:resource-baseline}) is unaware of our hardware slice abstraction, and the entire CGRA serves as a single large execution region.
Since an existing task is already mapped onto the CGRA, subsequent tasks are always forced to wait until the previous tasks finish and release the single execution region.

The simplest mechanism to form an execution region is only to support fixed-sized regions.
For example, all execution regions in Figure~\ref{fig:fixed} consist of two GLB-slices and one array-slice.
Fixed-sized regions are not optimal. Since each task must fit within the fixed-sized execution region, the largest task with the highest resource usage determines the size.
On the other hand, when there are several available execution regions, a task can be unrolled and mapped in parallel to achieve higher throughput (e.g., the next-running task is unrolled by three in Figure~\ref{fig:fixed}).
This method does not require much architectural change, and the implementation of a scheduling algorithm can be straightforward given the assumption that all target tasks fit within an execution region.
% However, the size of an execution region must be carefully determined so as not to hurt resource utilization.
However, although unrolling increases throughput, optimization across the unrolled dimension can be challenging to support.

Another method is to support variably sized execution regions by merging multiple fixed-sized regions.
We define the unit size of a region as in the fixed-sized region case, but we can merge multiple unit regions to form a larger execution region.
For example, in Figure~\ref{fig:variable}, three unit-sized regions are merged to execute the next-running task (colored in red).
The benefit of variably sized execution regions is to allow compilation optimization across the unrolled dimension.
For example, a \textit{camera pipeline} task with 3 pixels/cycle throughput uses four array-slices (Table~\ref{tab:task-resource}).
Naively unrolling it by four achieves 12 pixels/cycle throughput using 16 array-slices.
However, the compiler can optimize to time-multiplex PE tiles and achieve 12 pixels/cycle throughput with only six array-slices.
Support for a variably sized region still allows for the pre-computation of bitstreams for multiple variants of tasks with different resource usage and throughput.
However, this approach may still suffer from low resource utilization since the ratio of GLB-slices and array-slices within an execution region always remains the same.

Therefore, we propose \textit{flexible-shape execution regions} in which GLB-slices and array-slices are no longer coupled.
Decoupling of GLB-slices and array-slices enables finer-grained resource allocation.
For example, Figure~\ref{fig:flexible} shows how an execution region can be allocated any number of GLB-slices and array-slices, forming a non-rectangular shape, with remaining array-slices and GLB-slices available to be used by other tasks.
The support for flexible-shape execution regions improves resource utilization, especially for multi-tasked workloads where memory-intensive and compute-intensive tasks are mixed.
However, it may require additional communication between the GLB-slices and the array-slices.
In this work, we limit the placement of GLB-slices and array-slices within an execution region to be contiguous to simplify our study.
Design space exploration on flexible placement support and the required network remains as future work.
Section~\ref{subsec:cloud} describes the benefits of these mechanisms in more detail with a cloud system example.

% \begin{figure}
%     \centering
%     \includegraphics[width=1.0\linewidth]{example-image}
%     \caption{A multi-stage network.}
%     \label{fig:network}
% \end{figure}

% \medskip\noindent\textbf{Multi-Stage Network.}

% Figure~\ref{fig:network} shows
% when reconfiguration needed, stall.

% \begin{figure}
%     \centering
%     \includegraphics[width=1.0\linewidth]{example-image}
%     \caption{DPR}
%     \label{fig:dpr}
% \end{figure}

\noindent\textbf{Dynamic Partial Reconfiguration.}
Dynamic partial reconfiguration (DPR) is a mechanism to update the hardware configuration in reconfigurable architectures.
We propose fast-DPR following the DPR mechanism proposed in Amber SoC~\cite{amber}, but with added features to exploit hardware abstractions.
% We propose fast-DPR where reconfiguration of the entire tile-array can be done in 7.0{\micro}s.
In Amber, every other GLB bank stores the configuration bitstreams and independently streams configuration into two columns of the tile array.
Also, clocks and configuration signals are distributed down each column together, enabling reconfiguring the tile array at high clock frequency without pipeline stages.
In our CGRA, we also reuse GLB blocks to store and stream bitstreams to the tile array and follow the same clock distribution network.
Unlike Amber, however, one GLB bank streams configuration into one array-slice (in turn, four columns of the tile array) as an array-slice is the minimum unit of execution regions.
% Specifically, as every set of four columns in the tile array consisting an array-slice, a GLB bank streams configuration into four columns.
% As there are eight array-slices in the CGRA, we can reconfigure each array-slice in parallel by using eight GLB banks.

We added a feature to relocate bitstreams at run time to exploit hardware abstractions further.
In Amber, the compiler generates region-aware bitstreams; the bitstreams for one region cannot be reused in different regions even though the two regions are homogeneous.
This limitation comes from the fact that the address of each configuration register in different columns has a distinct column \#id.
On the other hand, our compiler generates region-agnostic bitstreams by assuming that the task is always mapped to the leftmost region.
We also added a register indicating the destination region of DPR to GLB banks.
When the host processor triggers DPR, GLB banks read the register and stream bitstreams to the target region via the network between the GLB and the tile array.
With this bitstream relocation feature, a user can pre-load bitstreams of the next task to the GLB in advance and rapidly map it to any next available region just by writing to a single register.

% dpr overview. reconfiguration. number of registers. time.
% Figure~\ref{fig:dpr-network}

% \medskip\noindent\textbf{Clock Distribution.}
% Figure~\ref{fig:dpr-clock}

\section{Evaluation}
\label{sec:evaluation}

\begin{figure}
    \centering
    \subfloat[Cloud system example]{\includegraphics[width=0.43\textwidth]{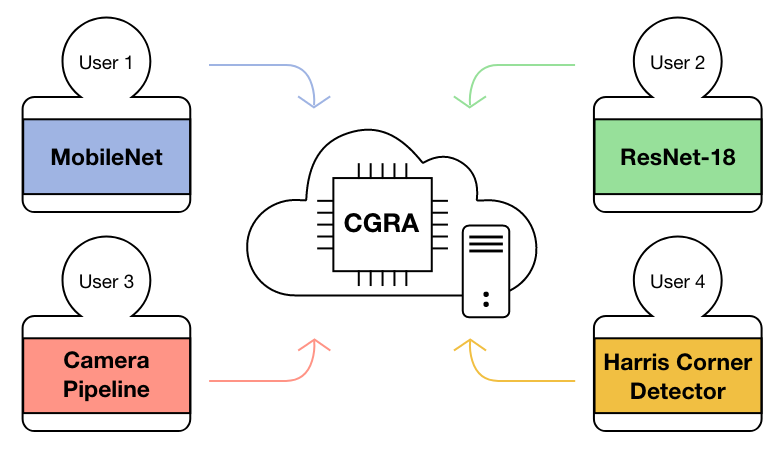}
    \label{fig:cloud}}\hskip1ex
    \subfloat[Autonomous system example]{\includegraphics[width=0.43\textwidth]{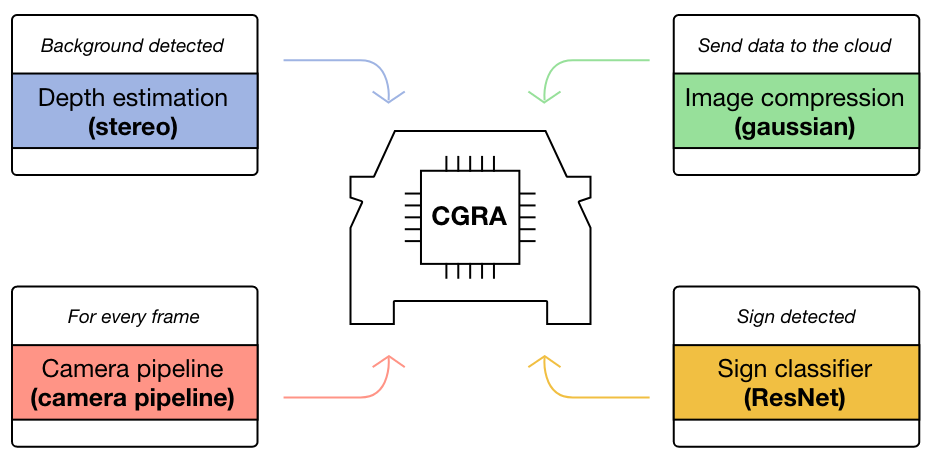}
    \label{fig:edge}}
    \caption{(a) Cloud system example scenario with four tenants submitting requests to the CGRA. Each tenant is assigned with a task from \textit{MobileNet}, \textit{ResNet-18}, \textit{camera pipeline}, and \textit{Harris}, respectively. (b) Autonomous system example with tasks that may be triggered under conditions.}
    \label{fig:example}
\end{figure}

We evaluate the benefits of multi-task execution support under two different workload scenarios.
In a cloud system example scenario (Section~\ref{subsec:cloud}), our CGRA with flexible-shape execution regions enables 1.05x-1.24x higher throughput and 23-28\% lower normalized turn-around time (NTAT) over the baseline CGRA.
In an autonomous system example scenario (Section~\ref{subsec:autonomous}), our CGRA enables 60.8\% reduced total latency.

\subsection{Example 1: Cloud System}
\label{subsec:cloud}
\noindent\textbf{Overview}. 
In this example, we construct a synthetic cloud computing scenario that models real-world examples in which the CGRA serves application requests from multiple users (Figure~\ref{fig:cloud}).
We construct the multi-tasked workload using kernels from machine learning (ML) and image processing domains, including ResNet-18~\cite{resnet} and MobileNet~\cite{mobilenet} from the ML domain, and camera pipeline and Harris corner detector from the image processing domain.
Table~\ref{tab:task-resource} summarizes the benchmark tasks and their resource requirements.

To generate the multi-tasked workload, we assume four tenants share the CGRA and are assigned one of the four target applications.
Each tenant sends a request to the CGRA following a Poisson distribution.
Whenever a new task arrives, or an existing task finishes, the scheduler is triggered and runs a greedy algorithm to schedule the next available task.
The scheduler checks if dependencies are met before scheduling the task (e.g., in ResNet-18, \textit{conv2\_x} depends on \textit{conv1\_x}).
If there is more than one version of a task that can be mapped onto the available resources, the greedy scheduler always chooses the one with the highest throughput.

\noindent\textbf{Metrics}.
We measure \textit{Normalized Turn-Around Time} and \textit{throughput} to compare the baseline CGRA and the three partitioning mechanisms described in Section~\ref{sec:partition}.
\textit{Turn-Around Time} (TAT) is the interval from the time of request to submit a task to the time of task completion.
\textit{Normalized Turn-Around Time} (NTAT) is the ratio of the TAT to the execution time, which represents the relative delay of a task (Equation (\ref{equ:tat}) - (\ref{equ:ntat})).
We calculate NTAT for each request and the arithmetic average for each application.
We also measure the average throughput for each application to demonstrate the performance benefit.
\setlength{\abovedisplayskip}{2.4pt}
\setlength{\belowdisplayskip}{-1.5pt}
\begin{equation}
\label{equ:tat}
TAT\; =\; \textit{wait\_time}\; +\; \textit{execution\_time}
\end{equation}
\begin{equation}
\label{equ:ntat}
NTAT\; =\; TAT\; /\; \textit{execution\_time}
\end{equation}

\begin{figure}
    \centering
    \subfloat[NTAT]{\includegraphics[width=0.43\textwidth]{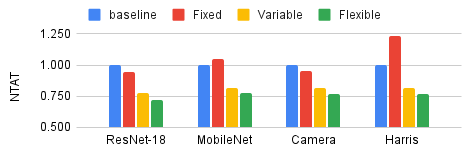}
    \label{fig:cloud-antt}}\hskip1ex
    \subfloat[Throughput]{\includegraphics[width=0.43\textwidth]{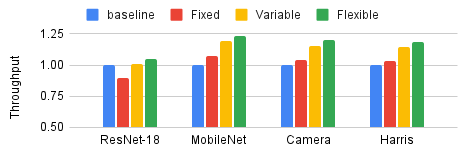}
    \label{fig:cloud-stp}}
    \caption{Evaluation in a cloud system example. (a) NTAT and (b) throughput for each task with fixed-sized, variably sized, and flexible-shape resource partitioning, normalized to the baseline CGRA. Flexible-shape partitioning decreases NTAT by 23-28\% and increases throughput by 1.05x-1.24x.}
    \label{fig:cloud-result}
\end{figure}
\noindent\textbf{Results}.
Figure~\ref{fig:cloud-result} illustrates the relative improvements in NTAT and throughput for flexible-shape execution regions compared to fixed- and variably-sized execution regions.
Even with a simple greedy scheduling algorithm, we achieve 23--28\% decreased NTAT and 1.05x--1.24x higher throughput.
Note that we only pre-compile each task to two different variants in this case study (Table~\ref{tab:task-resource}), and a scheduler greedily selects the one with higher throughput if resources are available.
Co-optimizing compilation and scheduling policy may improve NTAT and throughput further, which remains future work.

\subsection{Example 2: Autonomous System}
\label{subsec:autonomous}
\noindent\textbf{Overview}.
In this case study, we construct a synthetic edge system scenario modeling the real world in which multiple tasks from image processing and ML domains execute in parallel and can dynamically trigger.
Specifically, we develop an autonomous system scenario as described in Figure~\ref{fig:edge} following a methodology used in \cite{quantifying}.
\footnote{We also changed the tasks to simplify the example.}
The system takes a RAW image in Bayer encoding format (RGGB) from sensors at 30 fps and first runs a \textit{camera pipeline} task on the CGRA to convert to an RGB image.
Once the CGRA generates an RGB image, the system runs object detection and dynamically decides on the next tasks.
\footnote{This work assumes that object detection is executed in another hardware in the system (e.g. GPU or ASIC).}
When an event happens (e.g., detection of a specific background), it processes the event and executes the corresponding tasks (e.g., \textit{depth estimation}).
Except for a \textit{camera pipeline} that runs every frame, we set the period from one event to the next same event to follow a uniform random distribution between 3--7 frames.

\begin{figure}
    \centering
    \includegraphics[width=0.9\linewidth]{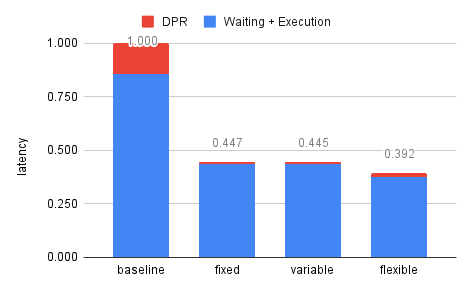}
    \caption{The average latency of an autonomous system example with different execution regions. The values are normalized to the result of the baseline.
    A red bar indicates the time spent for reconfiguration, and a blue bar indicates the sum of wait time and execution time. To show the benefit of fast-DPR (Section~\ref{sec:partition}), we assume the baseline CGRA uses AXI4-Lite interface for DPR, while others use fast-DPR.
    }
    \label{fig:edge-result}
\end{figure}

\noindent\textbf{Results}.
We evaluate the benefit of hardware resource partitioning and fast DPR by comparing our proposed CGRA to the baseline CGRA with AXI4-Lite-based DPR.
Specifically, the baseline CGRA maps only one task at a time.
When more than one event occurs, the baseline handles each task one by one and reconfigures using sequential AXI4-Lite configuration transactions.
In the proposed CGRA with multi-task execution support, we exploit flexible-shape resource partitioning to concurrently run more than one task on the CGRA when possible.
Also, we use the parallel and high-frequency DPR mechanisms in Section~\ref{sec:partition} to configure bitstreams.
We compute the arithmetic average of the latency over all frames.
As described in Figure~\ref{fig:edge-result}, our techniques enable a 60.8\% latency reduction compared to the baseline.
With fast DPR, reconfiguration takes less than 5\% of the total latency, an appreciable reduction from 14.4\% in the baseline.

% \begin{figure}
%     \centering
%     \subfloat[Latency]{\includegraphics[width=0.47\textwidth]{figs/latency.png}
%     \label{fig:latency}}\hskip1ex
%     \subfloat[Energy]{\includegraphics[width=0.47\textwidth,height=3cm]{example-image}
%     \label{fig:energy}}
%     \caption{(a) Latency (b) Energy}
%     \label{fig:edge-result}
% \end{figure}

\section{Related Work}
\label{sec:related_work}

As Deep Neural Networks (DNNs) are widely used in various domains, DNN accelerators~\cite{eie, ese, eyeriss, eyeriss2, dadiannao, maeri} have emerged and been deployed in the cloud system~\cite{tpu, brainwave}.
To that end, many prior works have explored multi-tenancy support on DNN accelerators in cloud systems.
Multi-task execution support is also studied in FPGAs targeting both cloud and edge computing.
However, a non-negligible portion of FPGA resources is typically reserved for controlling multi-task execution, ultimately decreasing the available computing resources.
ChordMap~\cite{chordmap} explores the automated mapping of multi-tasked applications onto a CGRA, but it is limited to mapping multiple tasks within streaming applications with all tasks known a priori.
Our work proposes hardware abstractions and mechanisms, which both compilers and schedulers can exploit and co-optimize to improve resource utilization in both cloud and edge systems.

\noindent\textbf{Multi-Task Execution on DNN Accelerators}.
Some DNN accelerators service multi-DNN tasks at the software level.
AI-MT~\cite{aimt} and Layerweaver~\cite{layerweaver} propose a scheduling policy to mix compute- and memory-intensive tasks to increase hardware utilization.
PREMA~\cite{prema} implements preemptible NPUs to support multi-tenancy via temporal multiplexing.
Many works add flexibility to an accelerator to accommodate multiple DNN tasks.
Planaria~\cite{planaria} introduces a flexible systolic array with dynamic architecture fission to map multiple DNN tasks.
\cite{mirroring} suggests a multi-directional network to support up to four DNN tasks with different dataflow.
Other works~\cite{heterogeneous, google} explore a computing system with multiple DNN accelerators with different hardware characteristics.
While these works only support DNN workloads, our work can support any applications that can be mapped onto a CGRA. 

\noindent\textbf{Multi-Task Execution on FPGAs}.
In FPGAs, multi-task execution support has been explored in the context of virtualization.
Some works divide an FPGA into a static region, a \textit{shell}, which serves as glue logic between the host and the FPGA, and a dynamic region, a \textit{role}, which handles the computation of tasks.
\cite{fpga1, fpga2, fpga3} partition a physical FPGA into several fixed-size virtual blocks and share them across multiple tasks.
AmorphOS~\cite{amorphos} presents a hardware abstraction of an FPGA, ~\textit{Morphlet}, which dynamically alters its size based on resource requirements.
ViTAL~\cite{vital} provides a full-stack framework to run multiple tasks with different sizes on homogeneous regions.
\cite{deepfpga} supports running multi-DNN tasks on an FPGA by dividing hardware resources into multiple PE cores and spatially multiplexing them, while ~\cite{quantifying} evaluates the benefits of temporal multiplexing of FPGAs using DPR for vision applications on embedded devices.
While these works only target scenarios where underlying applications change infrequently because of long reconfiguration time of FPGAs, our work can support both cloud systems and real-time edge systems due to rapid DPR.
% However, our work supports fast dynamic partial reconfiguration, which makes it usable for real-time applications as well.

\section{Conclusion}
\label{sec:conclusion}
Multi-task execution support on accelerators is becoming increasingly relevant in both cloud and edge systems and has the potential to improve performance through better hardware utilization.
This work proposes abstracting hardware resources within a CGRA into coarser-grained units with which a workload scheduler can quickly make decisions.
Based on the proposed abstraction, we develop hardware mechanisms to support multi-task execution through flexible-shape hardware partitioning and high-throughput dynamic partial reconfiguration.
Our evaluations modeling both a cloud and an edge system scenario suggest that the abstraction and hardware mechanisms can enable automatic schedulers to achieve high performance in multi-tasked workloads on future CGRAs.

%%%%%%% -- PAPER CONTENT ENDS -- %%%%%%%%

\bibliographystyle{IEEEtranS}
\bibliography{references}

% Generated by IEEEtranS.bst, version: 1.14 (2015/08/26)
\begin{thebibliography}{10}
\providecommand{\url}[1]{#1}
\csname url@samestyle\endcsname
\providecommand{\newblock}{\relax}
\providecommand{\bibinfo}[2]{#2}
\providecommand{\BIBentrySTDinterwordspacing}{\spaceskip=0pt\relax}
\providecommand{\BIBentryALTinterwordstretchfactor}{4}
\providecommand{\BIBentryALTinterwordspacing}{\spaceskip=\fontdimen2\font plus
\BIBentryALTinterwordstretchfactor\fontdimen3\font minus
  \fontdimen4\font\relax}
\providecommand{\BIBforeignlanguage}[2]{{%
\expandafter\ifx\csname l@#1\endcsname\relax
\typeout{** WARNING: IEEEtranS.bst: No hyphenation pattern has been}%
\typeout{** loaded for the language `#1'. Using the pattern for}%
\typeout{** the default language instead.}%
\else
\language=\csname l@#1\endcsname
\fi
#2}}
\providecommand{\BIBdecl}{\relax}
\BIBdecl

\bibitem{egra}
G.~Ansaloni, P.~Bonzini, and L.~Pozzi, ``Egra: A coarse grained reconfigurable
  architectural template,'' \emph{IEEE Transactions on Very Large Scale
  Integration (VLSI) Systems}, vol.~19, no.~6, pp. 1062--1074, 2010.

\bibitem{aimt}
E.~Baek, D.~Kwon, and J.~Kim, ``A multi-neural network acceleration
  architecture,'' in \emph{2020 ACM/IEEE 47th Annual International Symposium on
  Computer Architecture (ISCA)}.\hskip 1em plus 0.5em minus 0.4em\relax IEEE,
  2020, pp. 940--953.

\bibitem{google}
A.~Boroumand, S.~Ghose, B.~Akin, R.~Narayanaswami, G.~F. Oliveira, X.~Ma,
  E.~Shiu, and O.~Mutlu, ``Google neural network models for edge devices:
  Analyzing and mitigating machine learning inference bottlenecks,'' in
  \emph{2021 30th International Conference on Parallel Architectures and
  Compilation Techniques (PACT)}.\hskip 1em plus 0.5em minus 0.4em\relax IEEE,
  2021, pp. 159--172.

\bibitem{fpga1}
S.~Byma, J.~G. Steffan, H.~Bannazadeh, A.~Leon-Garcia, and P.~Chow, ``Fpgas in
  the cloud: Booting virtualized hardware accelerators with openstack,'' in
  \emph{2014 IEEE 22nd Annual International Symposium on Field-Programmable
  Custom Computing Machines}, 2014, pp. 109--116.

\bibitem{fpga2}
------, ``Fpgas in the cloud: Booting virtualized hardware accelerators with
  openstack,'' in \emph{2014 IEEE 22nd Annual International Symposium on
  Field-Programmable Custom Computing Machines}, 2014, pp. 109--116.

\bibitem{dream}
F.~Campi, A.~Deledda, M.~Pizzotti, L.~Ciccarelli, P.~Rolandi, C.~Mucci,
  A.~Lodi, A.~Vitkovski, and L.~Vanzolini, ``A dynamically adaptive dsp for
  heterogeneous reconfigurable platforms,'' in \emph{2007 Design, Automation \&
  Test in Europe Conference \& Exhibition}.\hskip 1em plus 0.5em minus
  0.4em\relax IEEE, 2007, pp. 1--6.

\bibitem{amber}
A.~Carsello, K.~Feng, T.~Kong, K.~Koul, Q.~Liu, J.~Melchert, G.~Nyengele,
  M.~Strange, K.~Zhang, A.~Nayak \emph{et~al.}, ``Amber: A 367 gops, 538 gops/w
  16nm soc with a coarse-grained reconfigurable array for flexible acceleration
  of dense linear algebra,'' in \emph{2022 IEEE Symposium on VLSI Technology
  and Circuits (VLSI Technology and Circuits)}.\hskip 1em plus 0.5em minus
  0.4em\relax IEEE, 2022, pp. 70--71.

\bibitem{eyeriss}
Y.-H. Chen, J.~Emer, and V.~Sze, ``Eyeriss: A spatial architecture for
  energy-efficient dataflow for convolutional neural networks,'' \emph{ACM
  SIGARCH Computer Architecture News}, vol.~44, no.~3, pp. 367--379, 2016.

\bibitem{eyeriss2}
Y.-H. Chen, T.-J. Yang, J.~Emer, and V.~Sze, ``Eyeriss v2: A flexible
  accelerator for emerging deep neural networks on mobile devices,'' \emph{IEEE
  Journal on Emerging and Selected Topics in Circuits and Systems}, vol.~9,
  no.~2, pp. 292--308, 2019.

\bibitem{dadiannao}
Y.~Chen, T.~Luo, S.~Liu, S.~Zhang, L.~He, J.~Wang, L.~Li, T.~Chen, Z.~Xu,
  N.~Sun \emph{et~al.}, ``Dadiannao: A machine-learning supercomputer,'' in
  \emph{2014 47th Annual IEEE/ACM International Symposium on
  Microarchitecture}.\hskip 1em plus 0.5em minus 0.4em\relax IEEE, 2014, pp.
  609--622.

\bibitem{prema}
Y.~Choi and M.~Rhu, ``Prema: A predictive multi-task scheduling algorithm for
  preemptible neural processing units,'' in \emph{2020 IEEE International
  Symposium on High Performance Computer Architecture (HPCA)}.\hskip 1em plus
  0.5em minus 0.4em\relax IEEE, 2020, pp. 220--233.

\bibitem{msftfpga}
J.~Fowers, K.~Ovtcharov, M.~Papamichael, T.~Massengill, M.~Liu, D.~Lo,
  S.~Alkalay, M.~Haselman, L.~Adams, M.~Ghandi \emph{et~al.}, ``A configurable
  cloud-scale dnn processor for real-time ai,'' in \emph{2018 ACM/IEEE 45th
  Annual International Symposium on Computer Architecture (ISCA)}.\hskip 1em
  plus 0.5em minus 0.4em\relax IEEE, 2018, pp. 1--14.

\bibitem{brainwave}
------, ``A configurable cloud-scale dnn processor for real-time ai,'' in
  \emph{2018 ACM/IEEE 45th Annual International Symposium on Computer
  Architecture (ISCA)}.\hskip 1em plus 0.5em minus 0.4em\relax IEEE, 2018, pp.
  1--14.

\bibitem{planaria}
S.~Ghodrati, B.~H. Ahn, J.~Kyung~Kim, S.~Kinzer, B.~R. Yatham, N.~Alla,
  H.~Sharma, M.~Alian, E.~Ebrahimi, N.~S. Kim, C.~Young, and H.~Esmaeilzadeh,
  ``Planaria: Dynamic architecture fission for spatial multi-tenant
  acceleration of deep neural networks,'' in \emph{2020 53rd Annual IEEE/ACM
  International Symposium on Microarchitecture (MICRO)}, 2020, pp. 681--697.

\bibitem{snafu}
G.~Gobieski, A.~O. Atli, K.~Mai, B.~Lucia, and N.~Beckmann, ``Snafu: an
  ultra-low-power, energy-minimal cgra-generation framework and architecture,''
  in \emph{2021 ACM/IEEE 48th Annual International Symposium on Computer
  Architecture (ISCA)}.\hskip 1em plus 0.5em minus 0.4em\relax IEEE, 2021, pp.
  1027--1040.

\bibitem{angeleye}
K.~Guo, L.~Sui, J.~Qiu, J.~Yu, J.~Wang, S.~Yao, S.~Han, Y.~Wang, and H.~Yang,
  ``Angel-eye: A complete design flow for mapping cnn onto embedded fpga,''
  \emph{IEEE transactions on computer-aided design of integrated circuits and
  systems}, vol.~37, no.~1, pp. 35--47, 2017.

\bibitem{ese}
S.~Han, J.~Kang, H.~Mao, Y.~Hu, X.~Li, Y.~Li, D.~Xie, H.~Luo, S.~Yao, Y.~Wang
  \emph{et~al.}, ``Ese: Efficient speech recognition engine with sparse lstm on
  fpga,'' in \emph{Proceedings of the 2017 ACM/SIGDA International Symposium on
  Field-Programmable Gate Arrays}, 2017, pp. 75--84.

\bibitem{eie}
S.~Han, X.~Liu, H.~Mao, J.~Pu, A.~Pedram, M.~A. Horowitz, and W.~J. Dally,
  ``Eie: Efficient inference engine on compressed deep neural network,''
  \emph{ACM SIGARCH Computer Architecture News}, vol.~44, no.~3, pp. 243--254,
  2016.

\bibitem{resnet}
K.~He, X.~Zhang, S.~Ren, and J.~Sun, ``Deep residual learning for image
  recognition,'' in \emph{Proceedings of the IEEE conference on computer vision
  and pattern recognition}, 2016, pp. 770--778.

\bibitem{mobilenet}
A.~G. Howard, M.~Zhu, B.~Chen, D.~Kalenichenko, W.~Wang, T.~Weyand,
  M.~Andreetto, and H.~Adam, ``Mobilenets: Efficient convolutional neural
  networks for mobile vision applications,'' \emph{arXiv preprint
  arXiv:1704.04861}, 2017.

\bibitem{tpu}
N.~P. Jouppi, C.~Young, N.~Patil, D.~Patterson, G.~Agrawal, R.~Bajwa, S.~Bates,
  S.~Bhatia, N.~Boden, A.~Borchers \emph{et~al.}, ``In-datacenter performance
  analysis of a tensor processing unit,'' in \emph{Proceedings of the 44th
  annual international symposium on computer architecture}, 2017, pp. 1--12.

\bibitem{amorphos}
A.~Khawaja, J.~Landgraf, R.~Prakash, M.~Wei, E.~Schkufza, and C.~J. Rossbach,
  ``Sharing, protection, and compatibility for reconfigurable fabric with
  $\{$AmorphOS$\}$,'' in \emph{13th USENIX Symposium on Operating Systems
  Design and Implementation (OSDI 18)}, 2018, pp. 107--127.

\bibitem{ahatecs}
\BIBentryALTinterwordspacing
K.~Koul, J.~Melchert, K.~Sreedhar, L.~Truong, G.~Nyengele, K.~Zhang, Q.~Liu,
  J.~Setter, P.-H. Chen, Y.~Mei, M.~Strange, R.~Daly, C.~Donovick, A.~Carsello,
  T.~Kong, K.~Feng, D.~Huff, A.~Nayak, R.~Setaluri, J.~Thomas, N.~Bhagdikar,
  D.~Durst, Z.~Myers, N.~Tsiskaridze, S.~Richardson, R.~Bahr, K.~Fatahalian,
  P.~Hanrahan, C.~Barrett, M.~Horowitz, C.~Torng, F.~Kjolstad, and P.~Raina,
  ``Aha: An agile approach to the design of coarse-grained reconfigurable
  accelerators and compilers,'' \emph{ACM Trans. Embed. Comput. Syst.}, apr
  2022, just Accepted. [Online]. Available:
  \url{https://doi.org/10.1145/3534933}
\BIBentrySTDinterwordspacing

\bibitem{heterogeneous}
H.~Kwon, L.~Lai, M.~Pellauer, T.~Krishna, Y.-H. Chen, and V.~Chandra,
  ``Heterogeneous dataflow accelerators for multi-dnn workloads,'' in
  \emph{2021 IEEE International Symposium on High-Performance Computer
  Architecture (HPCA)}.\hskip 1em plus 0.5em minus 0.4em\relax IEEE, 2021, pp.
  71--83.

\bibitem{maeri}
H.~Kwon, A.~Samajdar, and T.~Krishna, ``Maeri: Enabling flexible dataflow
  mapping over dnn accelerators via reconfigurable interconnects,'' \emph{ACM
  SIGPLAN Notices}, vol.~53, no.~2, pp. 461--475, 2018.

\bibitem{mirroring}
J.~Lee, J.~Choi, J.~Kim, J.~Lee, and Y.~Kim, ``Dataflow mirroring:
  Architectural support for highly efficient fine-grained spatial multitasking
  on systolic-array npus,'' in \emph{2021 58th ACM/IEEE Design Automation
  Conference (DAC)}.\hskip 1em plus 0.5em minus 0.4em\relax IEEE, 2021, pp.
  247--252.

\bibitem{chordmap}
Z.~Li, D.~Wijerathne, X.~Chen, A.~Pathania, and T.~Mitra, ``Chordmap: Automated
  mapping of streaming applications onto cgra,'' \emph{IEEE Transactions on
  Computer-Aided Design of Integrated Circuits and Systems}, vol.~41, no.~2,
  pp. 306--319, 2022.

\bibitem{cgravideo}
L.~Liu, D.~Wang, M.~Zhu, Y.~Wang, S.~Yin, P.~Cao, J.~Yang, and S.~Wei, ``An
  energy-efficient coarse-grained reconfigurable processing unit for
  multiple-standard video decoding,'' \emph{IEEE Transactions on Multimedia},
  vol.~17, no.~10, pp. 1706--1720, 2015.

\bibitem{fpgavirt}
J.~Mbongue, F.~Hategekimana, D.~T. Kwadjo, D.~Andrews, and C.~Bobda,
  ``Fpgavirt: A novel virtualization framework for fpgas in the cloud,'' in
  \emph{2018 IEEE 11th International Conference on Cloud Computing
  (CLOUD)}.\hskip 1em plus 0.5em minus 0.4em\relax IEEE, 2018, pp. 862--865.

\bibitem{quantifying}
M.~Nguyen, R.~Tamburo, S.~Narasimhan, and J.~C. Hoe, ``Quantifying the benefits
  of dynamic partial reconfiguration for embedded vision applications,'' in
  \emph{2019 29th International Conference on Field Programmable Logic and
  Applications (FPL)}.\hskip 1em plus 0.5em minus 0.4em\relax IEEE, 2019, pp.
  129--135.

\bibitem{layerweaver}
Y.~H. Oh, S.~Kim, Y.~Jin, S.~Son, J.~Bae, J.~Lee, Y.~Park, D.~U. Kim, T.~J.
  Ham, and J.~W. Lee, ``Layerweaver: Maximizing resource utilization of neural
  processing units via layer-wise scheduling,'' in \emph{2021 IEEE
  International Symposium on High-Performance Computer Architecture
  (HPCA)}.\hskip 1em plus 0.5em minus 0.4em\relax IEEE, 2021, pp. 584--597.

\bibitem{jade}
A.~Vasilyev, N.~Bhagdikar, A.~Pedram, S.~Richardson, S.~Kvatinsky, and
  M.~Horowitz, ``Evaluating programmable architectures for imaging and vision
  applications,'' in \emph{2016 49th Annual IEEE/ACM International Symposium on
  Microarchitecture (MICRO)}.\hskip 1em plus 0.5em minus 0.4em\relax IEEE,
  2016, pp. 1--13.

\bibitem{fpga3}
J.~Weerasinghe, F.~Abel, C.~Hagleitner, and A.~Herkersdorf, ``Enabling fpgas in
  hyperscale data centers,'' in \emph{2015 IEEE 12th Intl Conf on Ubiquitous
  Intelligence and Computing and 2015 IEEE 12th Intl Conf on Autonomic and
  Trusted Computing and 2015 IEEE 15th Intl Conf on Scalable Computing and
  Communications and Its Associated Workshops (UIC-ATC-ScalCom)}, 2015, pp.
  1078--1086.

\bibitem{deepfpga}
S.~Zeng, G.~Dai, H.~Sun, K.~Zhong, G.~Ge, K.~Guo, Y.~Wang, and H.~Yang,
  ``Enabling efficient and flexible fpga virtualization for deep learning in
  the cloud,'' in \emph{2020 IEEE 28th Annual International Symposium on
  Field-Programmable Custom Computing Machines (FCCM)}.\hskip 1em plus 0.5em
  minus 0.4em\relax IEEE, 2020, pp. 102--110.

\bibitem{vital}
Y.~Zha and J.~Li, ``Virtualizing fpgas in the cloud,'' in \emph{Proceedings of
  the Twenty-Fifth International Conference on Architectural Support for
  Programming Languages and Operating Systems}, 2020, pp. 845--858.

\end{thebibliography}

\end{document}